An empirical approach to measuring interface energies in mixed-phase bismuth ferrite

Stuart R. Burns[*,†,#], Oliver Paull[†], Ralph Bulanadi[†,⊥], Christie Lau[†], Daniel Sando[*,†], J. Marty Gregg[‡], and Nagarajan Valanoor[†]

†School of Materials Science and Engineering, UNSW Sydney, NSW 2052, Australia

‡Centre for Nanostructured Media, School of Mathematics and Physics, Queen's University Belfast, University Road, Belfast BT7 1NN, UK

#Present address: Department of Chemistry, University of Calgary, 2500 University Drive NW, Calgary AB T2N 1N4, Canada

⊥Present address: Department of Quantum Matter Physics, University of Geneva, 24 Quai Ernest-Ansermet, CH-1211, Geneva 4, Switzerland

*stuart.burns@ucalgary.ca; daniel.sando@unsw.edu.au

KEYWORDS

bismuth ferrite, thin films, strain engineering, thickness effects, energetics

ABSTRACT

In complex oxide heteroepitaxy, strain engineering is a powerful tool to obtain phases in thin films that may be otherwise unstable in bulk. A successful example of this approach is mixed phase bismuth ferrite ($BiFeO_3$) epitaxial thin films. The coexistence of a tetragonal-like (T-like) matrix and rhombohedral-like (R-like) striations provides an enhanced electromechanical response, along with other attractive functional behaviors. In this paper, we compare the energetics associated with two thickness dependent strain relaxation mechanisms in this system: domain walls arising from monoclinic distortion in the T-like phase, and the interphase boundary between the host T-like matrix and tilted R-like phases. Combining x-ray diffraction measurements with scanning probe microscopy, we extract quantitative values using an empirical energy balance approach. The domain wall and phase boundary energies are found to be $113 \pm 21$ and $426 \pm 23$ $mJ.m^{-2}$, respectively. These numerical estimates will help us realize designer phase boundaries in multiferroics, which possess colossal responses to external stimuli, attractive for a diverse range of functional applications.

**INTRODUCTION**

Bismuth ferrite (BiFeO$_3$; BFO) is a multiferroic perovskite oxide which forms in the *R*3*c* space group, extensively studied for its room temperature ferroelectric, electromechanical, and antiferromagnetic properties [1,2]. Although BFO was first synthesized in epitaxial thin film form in 2003 (Ref. [3]), a dramatic surge in interest – particularly regarding the effects of epitaxial strain – occurred in 2009 upon the discovery of a giant axial ratio phase of BFO ("T-like" or "T' BFO", with *c/a* = 1.23), induced by large (> 4%) compressive misfit strains [4]. Under intermediary strain states, the material can also be grown with a mixed phase microstructure, *i.e.* partially forming bulk-like tilted phases (R-like; R') within a matrix comprising the T-like phase of BFO (Ref. [5]). This "mixed phase BFO" is commonly observed when the film is grown on lanthanum aluminate (LaAlO$_3$; LAO) substrates, to thicknesses above ~25 nm. In mixed phase BFO, electrical switching transitions between both structural (T' and R') and polar variants (+T' and –T') [6], doping [7], enhanced response from electromechanical switching [8–13], pinning mechanisms [14], optical effects [15,16], ferroelectric tunnel junctions [17], and deterministic control of phases [18,19] have been studied extensively. These films routinely display an increase in mixed phase population upon increasing film thickness, driven by strain relaxation. Interestingly, in some cases where defects are incorporated throughout the film volume, the formation of the mixed phase can be completely inhibited [20–22]. This complex equilibrium provides a wide scope of stimuli which can locally tailor the functionality of the system.

To accurately describe these stimuli and the associated energy landscape, we require a numerical handle of the physical parameters governing the strain relaxation process. Strain relief in mixed phase BFO is manifested through the formation of new interfaces, either domain walls (DWs) or interphase boundaries (**Fig. 1**) [23]. However, an explicit value of the energy density of either of these remains unknown. Having a knowledge of these energy values would help guide studies of new phase space in BFO and similar epitaxial thin film multiferroics.

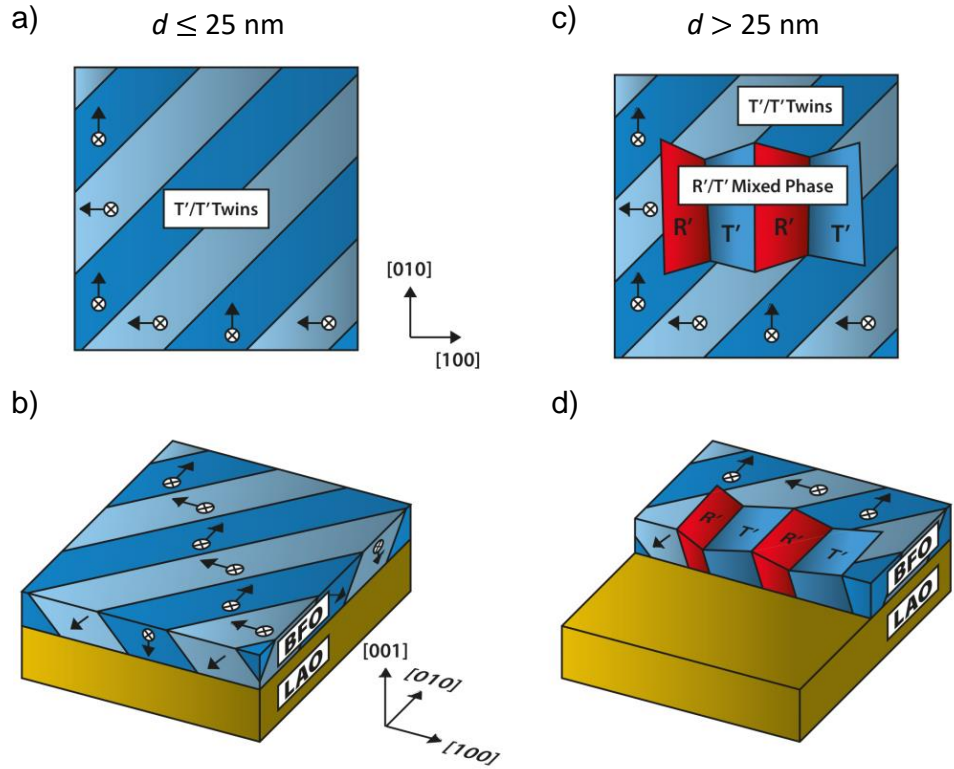

**Figure 1.** Schematic of various types of strain-relaxation mechanisms in strongly epitaxially-strained BFO films grown on LAO substrates. (a,b) only T'/T' twin DWs, which form along $\{1\bar{1}\bar{a}\}$ planes (where $a$ is related to the elastic compliances) [24], are observed for films with thickness below about 25 nm. (c,d) T'/T' twin DWs as well as T'/R' mixed phase boundaries are observed for films with thickness above about 25 nm. The arrows denote the local direction of ferroelectric polarization.

Here, we analyze the energy landscape associated with mixed phase BFO as a function of film thickness within the framework of an energy balance approach and a modification of the general Kittel's law [25]. Piezoresponse force microscopy (PFM) and atomic force microscopy (AFM) were used to calculate the periodicities of monoclinic domain walls within the T-like matrix (referred to as T'/T' twins) and the boundaries between coexistent structural phases (T'/R' phase boundaries), respectively. To obtain these periodicities, a substantial amount of data was collected and processed to ensure significant statistical weigh. X-ray diffraction (XRD) techniques were used to quantitatively demonstrate the thickness-dependent strain. Following thermodynamic analyses based on elastic potentials, we deduce values of the energy associated with each of these strain relief mechanisms.

## METHODS

**Thin film fabrication**

BiFeO$_3$ thin films were deposited on LaAlO$_3$ (001) substrates by pulsed laser deposition (PLD), using conditions reported previously [26]. A ceramic target of Bi$_{1.1}$FeO$_3$ was ablated at a repetition rate of 10 Hz using a KrF excimer laser (wavelength of 248 nm) in an oxygen partial pressure of ~ 0.1 Torr. The laser fluence and film growth rate were measured to be ~1.8 J cm$^{-2}$ and ~0.04 Å pulse$^{-1}$, respectively. The substrate was held at a temperature of 590 °C at a distance of ~ 5 cm from the target. After a predetermined number of pulses dependent on film thickness, each sample was cooled at a rate of 20 °C min$^{-1}$ in an oxygen pressure of 5 Torr.

**X-ray diffraction and structural characterization**

Structural characterization by x-ray diffraction, including high-angle *θ-2θ* scans and low-angle reflectometry (not shown), was carried out using K$_{α-1}$ radiation ($λ = 1.5406$ Å) on a Philips Materials Research Diffractometer (MRD), equipped with a two-bounce (220) Ge monochromator. To determine the phase composition (R' and T' phases) and the in- and out-of-plane lattice parameters, reciprocal space maps (RSMs) were collected on the same diffractometer using a 1D detector (PIXcel). The lattice parameters were calculated by two dimensional Gaussian peak-fitting the film peaks in the RSMs at each thickness. The uncertainty values for the lattice parameters were calculated from the uncertainty in the position of the peak centers of the 2D Gaussian fits, combined with the statistical uncertainty when calculating the lattice parameter from the various film spots from the two separate in-plane RSMs (namely 013 and 103).

**Atomic force microscopy and piezoresponse force microscopy**

Topography imaging and ferroic domain mapping were carried out on an Oxford Instruments Asylum Cypher in contact AFM and PFM modes. Due to the fact that T' BFO is monoclinic M$_C$ (Ref. [23]), with an in-plane component of polarization, the PFM data were collected with the lateral displacement channel. Vertical displacement (not shown) provided no contrast in the domain images in the regions of T'/T' walls. The AFM and PFM images were analyzed with WSxM software [27] through both line profiling and image flooding to calculate the periodicities of T'/T' and T'/R' interfaces.

**RESULTS AND DISCUSSION**

We study here a series of BFO films grown on LAO (001). This substrate induces a misfit strain of approximately -4.5%. As previously reported, the film grows in a pure T' phase at the growth temperature (590 °C), but during post-growth cooling to room temperature, the material relaxes to a mixture of tilted tetragonal-like and rhombohedral-like phases, of which the volume phase fractions are thickness dependent [28–30]. The films had thicknesses of 15-120 nm, as measured by x-ray reflectometry (XRR) for thinner films (thickness $d$ < 50 nm) (not shown), and the growth rate extrapolated to estimate the thicknesses for samples with $d$ > 50 nm. This sample set demonstrated the typically observed pure T-like phase in thinner films, while a mixed phase microstructure was present at higher thicknesses.

To establish a framework which can account for the two aforementioned strain relieving mechanisms, we must first calculate the strain energies and then periodicities for the T'/T' twins and the mixed phase interfaces. Thus, we begin with detailed XRD investigations to address the strain energies.

**Figure 2a, b)** present the results of XRD reciprocal space maps (RSMs) of the 120 nm-thick sample, around the 001 (symmetric) and 103 (asymmetric) reflections (the full maps for all film thicknesses are provided in **Figs. S1-S4**) [31]. The phases are labelled according to the following convention: T' is the tetragonal-like phase that forms the matrix of the film (the smooth, flat regions in **Fig. 1a**), T'$_{tilt}$ (R'$_{tilt}$) is the tilted T-like (secondary strained R-like) phase which is observed within the mixed-phase regions as striations on the film surface, and R'$_{relaxed}$ is the almost fully relaxed R-like phase (not observed in the AFM images). For the remainder of this paper, for simplicity we use T'/T' to denote the domain walls separating the T-like ferroelastic domains, and T'/R' to denote the interphase boundary between the T'$_{tilt}$ and R'$_{tilt}$ phases.

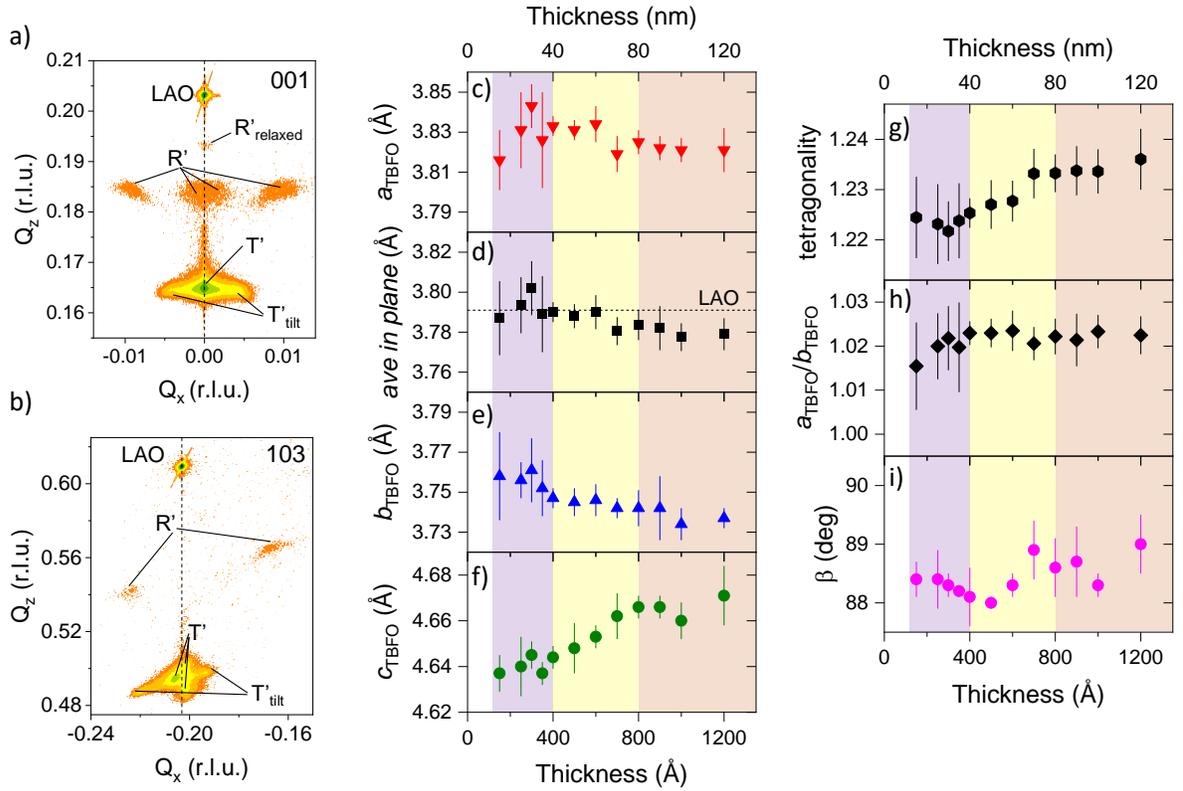

**Figure 2.** a) Symmetric x-ray diffraction RSM for the 120 nm-thick film near the (001) reflection of LAO and BFO. b) Asymmetric RSM near the (103) reflection for the 120 nm-thick film, analyzed to calculate in-plane lattice parameters. The three-fold splitting of the T' peak is a signature of a monoclinic $M_C$ crystal structure. The dashed lines in both RSMs indicate the theoretically calculated $Q_x$ value of the corresponding LAO reflection. c), d), e), f), g), h), i) $a$, average in-plane, $b$, and $c$ lattice parameters, tetragonality, $a/b$, and monoclinic angle $\beta$, of the T-like phase as a function of thickness, respectively. Note the position of the LAO pseudo-cubic lattice parameter (3.791 Å) with respect to the average in-plane parameters. Three thickness regimes are apparent, as discussed in the main text, and are shaded purple (~15 to ~40 nm), yellow (~40 to ~80 nm) and pink (~80 to ~120 nm).

In the XRD RSMs (**Fig. 2a, b**) several film diffraction peaks are observed, which correspond to the various phases related to the striations on the sample surface. We first consider the T' phase. As mentioned above, it is well established that T-like BFO typically forms as a $M_C$ monoclinic structure [29,32]. A signature of such crystallographic symmetry is the three-fold splitting of the T-like peak for the asymmetric (103) reflection, as labelled in **Fig. 2b)**. Combining symmetric and asymmetric RSMs measured on the full thickness series of films allows us to calculate we calculated the in-plane and out-of-plane lattice parameters of the T-like host matrix. These parameters are shown in **Fig. 2c), e), f)**, as well as the average in-plane lattice parameter $\frac{a+b}{2}$ (**Fig. 2d**), the tetragonality $\frac{2c}{a+b}$ (**Fig. 2g**), the distortion $a/b$ (**Fig. 2h**) and the monoclinic angle β (**Fig. 2i**).

Above 10-15 nm, the thickness-dependent behavior of the unit cell in the T-like phase can be separated into three regimes:

1) ~15 to ~40 nm (shaded in purple): The *a* and *b* lattice parameters for films in this thickness range, *i.e.*, prior to nucleation of mixed phase, clearly deviate from the pseudo-cubic lattice parameter of the LAO substrate (3.791 Å); however, the average in-plane parameter (**Fig. 2d**) is still consistent with that of the substrate. This phenomenon is considered to be a strain relaxation mechanism: instead of creating misfit dislocations to relieve strain, the film forms T'/T' domain walls. In this thickness range, the distortion increases with thickness (**Fig. 2h**). The out-of-plane lattice parameter, *c*, remains virtually unchanged (**Fig. 2f**).

2) ~40 to ~80 nm (shaded in yellow). For thickness values at which one would expect the mixed phase striations to form, the *c* lattice parameter increases, and the tetragonality of the unit cell also increases (**Fig. 2g**). The *a* and *b* parameter both decrease gradually.

3) ~80 to ~120 nm (shaded in pink). When a significant volume of mixed phase striations is formed in the thickest films, the *a* and *c* lattice parameters remain constant while the *b* parameter continues to decrease slightly. This is seemingly caused by the striations compressing the T-like unit cell. One could suggest that the freedom along the *b* axis in the T-like phase is linked to the preferential nucleation direction of the secondary phase.

We also used XRD RSMs to determine the lattice parameters of the relaxed R' phases which appear within the mixed phase striations, by combining the symmetric and asymmetric scans. This was only possible for films of thickness above 70 nm, since for thinner films, the XRD peaks for the secondary R' phase were too weak to obtain reliable parameters. Generally speaking, for the films 70-120 nm-thick, the lattice parameters of the R' phase tend only to marginally change for the thickest sample. The in-plane parameter slightly increases, while the out-of-plane parameter falls, with both migrating towards the BFO bulk pseudo-cubic lattice parameter [33] of 3.965 Å (**Fig. S5**) [31].

Having characterized the lattice parameters of each film, next we estimated the volume fractions of the R' and T' phases as a function of the thickness, particularly at the local scale. This thickness dependence of the mixed phase microstructure is more distinct when imaged through AFM. AFM imaging was carried out across the sample series in order to establish the ratio of mixed phase for a given film thickness (representative data for films 25-120 nm thick are given in **Figs. S6-S9**) [31]. Alongside this, an image processing technique was applied to quantify the density of the phase interfaces – boundaries between the stripe-like structures and their host matrix. **Figure S10** presents 5 x 5 μm² topography scans of sample surfaces for three film thicknesses [31], alongside images highlighting the approximate phase interfaces created with the microscopy analysis software described in Ref. [27].

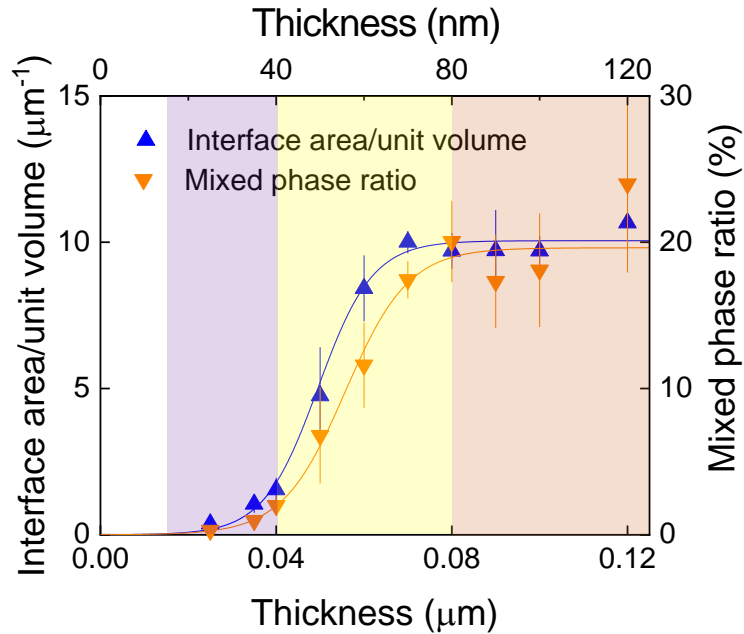

**Figure 3.** Average interface density (blue upward triangles) and R'-T' phase ratio (orange downward triangles) as a function of film thickness. The shading is in reference to the thickness regimes discussed in the main text and in Fig. 2. The solid lines are guides to the eye.

**Figure 3** presents the average R' phase percentage and interface density calculated and normalized to the film volume. Note that this is remarkably consistent with the thickness-dependent thermodynamic treatment of Ouyang and Roytburd [34,35]. The weighted average lattice parameter of the complete microstructure as a function of thickness is plotted in **Fig. S11** [31]. This was calculated by combining the data collected from XRD and AFM and shows the expected behavior of strain relaxation in the system – a decrease of the $c$ parameter and increase in average in-plane lattice parameter above a critical thickness of about 40 nm.

Next, we describe how the periodicities for both the T'/T' domain walls and the mixed phase interfaces were calculated. In T-like BFO, ferroelastic monoclinic domain structure has been previously reported [36–39]. These periodic walls, arising from the monoclinic distortion of the T-like unit cell, partially minimize elastic self-strain energy [40]. PFM was used to image this domain structure, with the amplitude shown in **Fig. 4a)** and **Fig. 4b)** for 40 and 80 nm thick films, respectively. In all cases the out-of-plane PFM phase was homogeneous, consistent with the out-of-plane polarization direction pointing downwards (not shown). Here, one can see the presence of tilted phase striations and ferroelastic DWs in the host T-like phase. All thicknesses studied here exhibited these ferroelastic DWs, implying that they are not as energetically costly as mixed phase interfaces.

These T'/T' domain walls are aligned along the <110>$_{pc}$ crystallographic directions (as expected for a $M_C$ symmetry) and terminate at the boundaries between the T-like regions and the mixed phase striations. The mixed phase striations themselves are tilted by up to 15 degrees from the high symmetry [100] or [010] directions (**Fig. 4b**) [41,42].

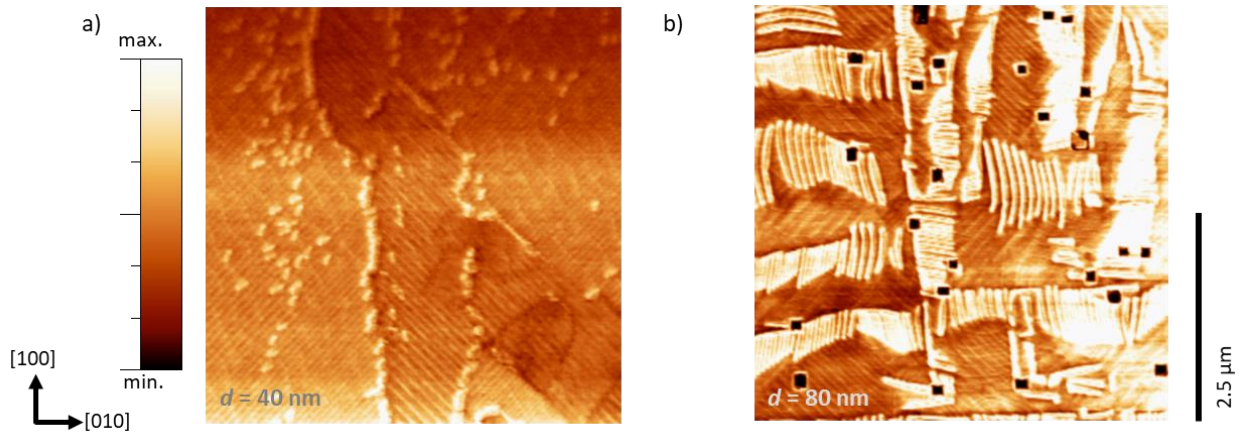

**Figure 4.** PFM amplitude images (5 x 5 μm$^2$) of; a) 40 nm-thick and b) 80 nm-thick films, where tilted phase structure and ferroelastic DWs are present.

Now that both the strain and periodicity for both types of interfaces have been obtained as a function of thickness, we proceed with the empirical analysis. For this, we start with the key expression employed in this work, Kittel's scaling law for ferroic domain walls [43], given in the general form as

$$w^2 = \frac{\gamma}{U}d \quad (1)$$

where $w$ is domain width, $\gamma$ is the domain wall energy, $U$ is the domain energy per unit volume, and $d$ is the film thickness. The form of $U$ is dependent on the ferroic order parameter associated with the domain. When empirically adopting this expression, two quantitative results must arise from experiments: 1) an energy term related to the order parameter and, 2) a periodicity of the discontinuity manifesting in the ferroic to reduce free energy. In this case, we exclusively consider elastic energy terms.

By modifying the general Kittel's law expression to add a *second* strain relieving mechanism, we can incorporate the role of phase interfaces in the analysis alongside the ferroelastic domain walls. As is typical for a Kittel expression, we start from free energy, $F$:

$$F = A_T \left( U_{T/T} w_{T/T} + \frac{\gamma_{T/T} d}{w_{T/T}} \right) + (1 - A_T) \left( U_{T/R} w_{T/R} + \frac{\gamma_{T/R} d}{w_{T/R}} \right) \quad (2)$$

where $A_T$ and $(1 - A_T)$ are the percentage fractions of T-like and mixed phase regions, respectively, $U_{T/T}$ is the elastic potential energy arising from the T-like monoclinic domain walls and $U_{T/R}$ is the elastic potential energy arising from the formation of mixed phase regions. The symbols $w$, $\gamma$ and $d$ have their usual meanings (given above), and the subscripts $T/T$ and $T/R$ denote which strain relieving interface the variables refer to T-like domain walls or mixed phase boundaries, respectively. Taking partial derivatives to describe energetic equilibrium, we arrive at the following relations:

$$\frac{\partial F}{\partial w_{T/T}} = A_T \left( U_{T/T} - \frac{\gamma_{T/T} d}{w_{T/T}^2} \right) = 0; \quad (3)$$

$$\frac{\partial F}{\partial w_{T/R}} = (1 - A_T) \left( U_{T/R} - \frac{\gamma_{T/R} d}{w_{T/R}^2} \right) = 0. \quad (4)$$

All of our samples exhibit T'/T' twins, therefore $A_T \neq 0$. When there is little to no mixed phase in the sample, $(1 - A_T) = 0$, and the equation reduces to the 'classic' Kittel's law, Equation (1). From this free energy expression, one can thus calculate the energy of formation for both the T'/T' twins ($\gamma_{T/T}$) and mixed phase microstructure ($\gamma_{T/R}$).

Returning to the experiments, we consider next the PFM and AFM derived periodicities for both T'/T' twins and mixed phase boundaries. **Figure 5** displays the periodicities of both strain relieving entities as a function of the square root of thickness (with intercept = 0), as a linear fitting of this plot would indicate that the scaling mechanism is following Kittel's law. As there are distinct thickness regions with different behaviors, we isolated the four thinnest films to calculate the periodicity of T'/T' twin walls. Using a similar line of reasoning, the four thickest films were isolated to calculate the periodicity of the phase boundaries. In this way, we identify two separate regimes to which we applied our energy analysis, each following a dense domain wall model. Rationale for using only these four end points in each case is provided later.

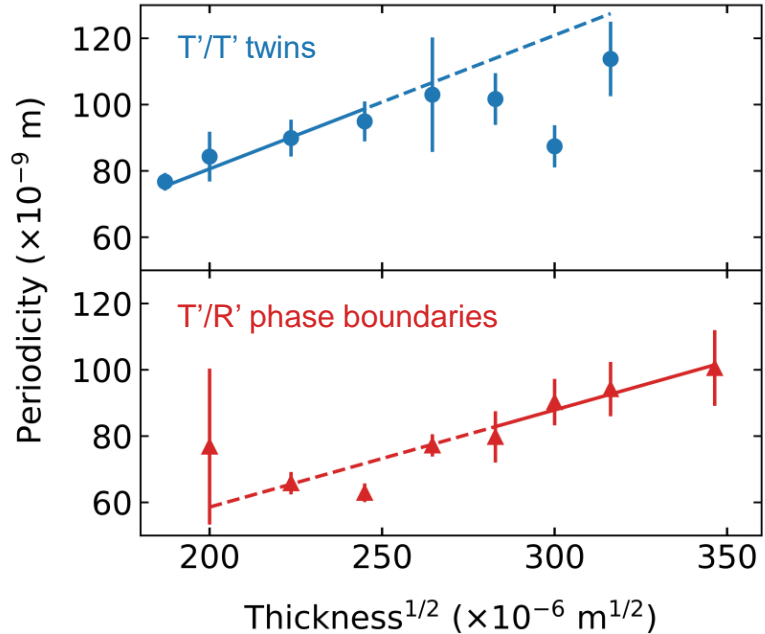

**Figure 5.** Periodicity against square root of film thickness of both T-like twin domain walls (top) and mixed phase boundaries (bottom) in the sample series of BFO//LAO (001). The linear fit for each data set indicates these strain relieving entities follow Kittel's scaling law (w∝$d^{1/2}$).

We denote the linear fit gradients as $M_{T/T}$ (blue fit, ≈ 4.03 x $10^{-4}$) and $M_{T/R}$ (red fit, ≈ 2.93 x $10^{-4}$) for the T-like domain walls and mixed phase boundaries, respectively. The gradients are related to the energy terms through the following equations, where $2\pi$ is a shape factor [44] typically used in Kittel expressions:

$$M_{T/T} = 2\pi \left(\frac{\gamma_{T/T}}{U_{T/T}}\right)^{1/2} \quad (5)$$

$$M_{T/R} = 2\pi \left(\frac{\gamma_{T/R}}{U_{T/R}}\right)^{1/2} \quad (6)$$

The elastic energy term $U_{T/T}$ was derived from the elastic misfit between the substrate and both the $a$ and $b$ lattice parameters:

$$\varepsilon_{a_T} = \frac{a_T - a_{sub}}{a_T} \quad (7)$$

$$\varepsilon_{b_T} = \frac{b_T - a_{sub}}{b_T} \quad (8)$$

$$U_{T/T} = \frac{Y}{1-\nu}\left(\varphi \varepsilon_{a_T}^2 + [1-\varphi]\varepsilon_{b_T}^2\right) \quad (9)$$

where $a_T$ and $b_T$ are the in-plane lattice parameters, $a_{sub}$ is the substrate lattice parameter of 3.791 Å, $Y$ is the Young's modulus (186 GPa, calculated from Ref. [34]), $\nu$ is the Poisson's ratio (0.35; Ref. [45]), and $\varphi$ is the phase fraction of $a$ oriented domains in the region of T-like phase, which, if assuming equal T'/T' domain populations, we take $\varphi = 1/2$.

Similarly, the energy $U_{T/R}$ was calculated by consideration of the average in-plane (IP) elastic misfit strains in the region of mixed phase striations:

$$\varepsilon_{IP_T} = \frac{IP_T - a_{sub}}{IP_T} \quad (10)$$

$$\varepsilon_{IP_R} = \frac{IP_R - a_{sub}}{IP_R} \quad (11)$$

$$U_{T/R} = \frac{Y}{1-\nu}\left(\varphi \varepsilon_{IP_T}^2 + [1-\varphi]\varepsilon_{IP_R}^2\right) \quad (12)$$

where $IP_T$ and $IP_R$ are the average in-plane lattice parameters for the T-like and R-like striations, and in this case $\varphi$ is the ratio of T-like to R-like phases in the areas containing striations, which we assume will reduce to $\varphi = 1/2$, given previous literature [28] identifying the phases of each portion of these needle-like structures.

The value of T'/T' domain wall (or phase boundary) energy, $\gamma_{T/T}$ (or $\gamma_{T/R}$) is extracted from the linear fits in **Fig. 5**:

$$\gamma_{T/T(T/R)} = \frac{U_{T/T(T/R)} M_{T/T(T/R)}^2}{4\pi^2}. \quad (13)$$

From this analysis, the average energy for forming T'/T' twin domain walls is found to be 113 ± 21 mJ.m$^{-2}$, while for forming mixed phase T'/R' boundaries it is 426 ± 23 mJ.m$^{-2}$. These values, as expected, are consistent with the experimental observation of a critical thickness at which mixed phases form – only at this thickness does the film overcome the cost of forming phase boundaries to relieve strain, whereas the more energetically favorable method of partial strain relief (forming T'/T' twins) occurs at all the thicknesses (above ~15 nm) across our sample range. These values of domain wall and interphase boundary energy are also of the same order of various types of domain walls in PbTiO$_3$ and BiFeO$_3$, as reported in Table 1.

| Material and domain wall variant | Domain wall energy (mJ/m$^2$) | Reference |
|---|---|---|
| BaTiO$_3$ - 180° wall | 7.2 - 16.8 | [46] |
| PbTiO$_3$ - 90° wall | 29.4 - 35.2 | [46] |
| PbTiO$_3$ - 180° wall | 132 - 169 | [46] |
| BiFeO$_3$ - 71° wall | 363 - 436 | [47] |
| BiFeO$_3$ - 109° wall | 205 - 1811 | [47] |
| BiFeO$_3$ - 180° wall | 829 | [47] |
| BiFeO$_3$ - T'/T' wall | 113 ± 21 | This work |
| BiFeO$_3$ - T'/R' phase boundary | 426 ± 23 | This work |

Table 1. Reported values of domain wall energy for BaTiO$_3$, PbTiO$_3$ (first principles calculations), and BiFeO$_3$ (first principles calculations and this work).

**DISCUSSION AND CONCLUSIONS**

Our empirical approach to determine domain wall and interphase boundaries in the mixed-phase BFO system comes with some caveats, which we briefly discuss here. Recall that our treatment uses two key quantitative results from experiments: i) a strain energy term related to the order parameter, and ii) a periodicity discontinuity, which we ascribe to the system changing the type of interface nucleated in order to reduce the elastic energy. As there is no change of phase in the out-of-plane PFM signal with increasing thickness, we surmise that these domains do not form to minimize depolarization field. Therefore, we have disregarded the electrostatic energies and assume that the formation of T'/T' occurs to minimize the internal self-strain energy arising from the monoclinic distortion of the T-like unit cell. This explains why we exclusively treat the free energy in-elastic energy terms.

Next, we point out that we have not accounted for the curvature of the domain walls. This would require incorporating a gradient term (in the vicinity of the wall) in the strain energy expression, only possible experimentally by precise mapping of the atomic level displacements as a function of the distance from the wall. To the best of our knowledge, such information would only be accessible using aberration corrected transmission electron microscopy which is beyond the scope of this paper. Therefore, we only fitted the first four data points to the T'/T' twin walls as we are certain with only these four samples the strain state is governed only by the T'/T' twins and not the mixed phase formation. In the same vein, only the last four data points are fitted to extract the energy for the T'/R' interphase boundaries. It is only in these last four samples that we truly find the dense domain state, which allows us to reasonably disregard the gradient terms [34,48].

Finally, for simplicity, our approach does not take into account the presence of other possible domain arrangements, such as nanotwins which can form parallel to the substrate interface, reported by Pailloux *et al.* [49]. The laser used for PLD grown in that work was Nd:YAG, while for our samples we used an excimer laser. It has been shown that the strongly different growth conditions for the two ablation laser types results in BFO films with different crystallinity, mosaicity, and domain structures [50,51]. Similarly, it is possible that the Nd:YAG laser promotes the formation of such exotic nanotwin defects in BFO//LAO. To the best of our knowledge, excimer grown BFO//LAO samples have not been reported to show such nanotwin structures.

In summary, we have explored the energetics of formation of T'/T' and interphase boundaries in the mixed-phase BiFeO$_3$ thin film system. Combining XRD, AFM, and PFM techniques, we characterized a series of BFO//LAO (001) films with thicknesses of 15-120 nm. From the XRD data we extracted the lattice parameters for both the T-like and mixed phase regions, allowing the estimation of elastic strain energies. The periodicities of the T'/T' twin ferroelastic domain walls and mixed phase T'/R' striations were determined through PFM and AFM analyses, respectively. The energy of formation for domain walls and interphase boundaries were calculated to be $\gamma_{T/T} \approx$ 113 ± 21 mJ.m$^{-2}$ and $\gamma_{T/R} \approx$ 426 ± 23 mJ.m$^{-2}$, respectively. These results strengthen our understanding of these strain relieving microstructures and provide numerical guidelines for the engineering of new exotic phases in tailor-made and dimensionally-confined [52] multiferroic systems.

AUTHOR INFORMATION

**Corresponding Authors**

*stuart.burns@ucalgary.ca; daniel.sando@unsw.edu.au

**Present Addresses**

# Department of Chemistry, University of Calgary, 2500 University Drive NW, Calgary AB T2N 1N4, Canada

⊥ Department of Quantum Matter Physics, University of Geneva, 24 Quai Ernest-Ansermet, CH-1211, Geneva 4, Switzerland

**Author Contributions**

The manuscript was written through contributions of all authors. All authors have given approval to the final version of the manuscript.

**Notes**

The authors declare no competing financial interest.

ACKNOWLEDGMENTS

This research was partially supported by the Australian Research Council Centre of Excellence in Future Low-Energy Electronics Technologies (Project No. CE170100039) and funded by the Australian Government. D.S. and V.N. acknowledge the support of the Australian Research Council through Discovery Grants. S.R.B. acknowledges current funding from the Canada First Research Excellence Fund, and partial funding from the UNSW Science PhD Writing Scholarship. S.R.B. and O.P. thank AINSE Limited for providing financial assistance (Award - PGRA). The authors thank Ekhard Salje, Alina Schilling and Marios Hadjimichael for helpful discussions.